\documentclass[sigconf]{acmart}

\copyrightyear{2020}
\acmYear{2020}
\setcopyright{acmcopyright}\acmConference[CIKM '20]{Proceedings of the 29th ACM
International Conference on Information and Knowledge Management}{October
19--23, 2020}{Virtual Event, Ireland}
\acmBooktitle{Proceedings of the 29th ACM International Conference on Information
and Knowledge Management (CIKM '20), October 19--23, 2020, Virtual Event, Ireland}
\acmPrice{15.00}
\acmDOI{10.1145/3340531.3412163}
\acmISBN{978-1-4503-6859-9/20/10}

\begin{document}
\fancyhead{}
\title{Diversifying Multi-aspect Search Results Using Simpson's Diversity Index }
\author{Jianghong Zhou}

\affiliation{%
  \institution{Emory University}
  \streetaddress{201 Dowman Drive}
  \city{Atlanta}
  \state{United States}
  \postcode{Ga. 30322}
}
\email{jianghong.zhou@emory.edu}

\author{Eugene Agichtein}
\affiliation{%
  \institution{Emory University}
  \city{Atlanta}
  \country{United States}}
\email{eugene.agichtein@emory.edu}

\author{Surya Kallumadi}
\affiliation{%
  \institution{The Home Depot}
  \city{Atlanta}
  \country{United States}}
\email{surya@ksu.edu}

\begin{abstract}
In search and recommendation, diversifying the multi-aspect search results could help with reducing redundancy, and promoting results that might not be shown otherwise. Many previous methods have been proposed for this task.
However, previous methods do not explicitly consider the uniformity of the number of the items' classes, or \textit{evenness},
which could degrade the search and recommendation quality. To address this problem, we introduce a novel method by adapting the Simpson's Diversity Index from biology, which enables a more effective and efficient quadratic search result diversification algorithm. 
We also extend the method to balance the diversity between multiple aspects through weighted factors and further improve computational complexity by developing a fast approximation algorithm. We demonstrate the feasibility of the proposed method using the openly available Kaggle shoes competition dataset. Our experimental results show that our approach outperforms previous state of the art diversification methods, while reducing computational complexity.
\end{abstract}


\maketitle

\section{Introduction}
Diversifying search results is of vital importance in many search systems due to the ambiguity of search queries and the complexity of users' intents. In multi-aspect search, a diverse set of search results are more likely to satisfy the users' needs and improve the online shopping experience. Therefore, many approaches for diversifying results in information retrieval and recommender systems have been proposed \cite{zhou2020rlirank}, among which the algorithms based on DPPs (Determinantal Point Processes) are considered to be  state of the art methods. DPPs aim to balance diversity and relevance, and are based on the geometric properties of vectors associated with a set of items \cite{kulesza2012determinantal}. DPPs define the diversity of a set by the volume of a parallelepiped spanned by the vectors of this set. The volume is determined by the vectors' length and their cosine similarity, which can represent the relevance and similarity respectively. Specifically, DPPs compute the determinant of the product of the selected items' matrix and its transpose to identify a diverse subset of results. 

Diversity is characterised by two aspects \textbf{1) richness} and \textbf{2) evenness} \cite{simpson1949measurement}. Richness quantifies the number of different classes elements in a set, while evenness considers the uniformity of the distribution of these classes. 
However, DPPs only consider the richness aspect
\cite{xie2017uncorrelation}.
It narrows the definition of diversity, which may degrade actual diverse search results. Besides, evenness can affect richness, especially the item has multiple aspects \cite{stirling2001empirical}.

In addition to not considering ``evenness'', DPPs also suffer from high computation complexity, which is NP hard. In its greedy formulation, the computation complexity is $O(N^4)$, in which $N$ is the scale of search space. Therefore, many variations are developed to improve its computational efficiency \cite{chen2017improving}\cite{chen2018fast}\cite{kulesza2011k}. 

Due to the aforementioned difficulties, we consider another model to measure diversity, which is Simpson's diversity index \cite{simpson1949measurement}. It measures the probability that two items are chosen at random and independently from a set is the same. Both Shannon entropy and Simpson's diversity index measure richness and evenness. Notwithstanding, Simpson's diversity index is a pairwise approach. Through some transformations and improvements, we develop it to a binary quadratic program, with predicable approximation bounds.

We evaluate our approach on a standard benchmark dataset from Kaggle shoes' price competition (\url{https://www.kaggle.com/datafiniti/womens-shoes-prices}, \url{https://www.kaggle.com/datafiniti/mens-shoe-prices}). The experimental results show that the proposed method outperforms state of the art DPP algorithms and extensions. In summary, our contributions are threefold:

\textbf{1.} Introducing Simpson's diversity index (SDI) to search result diversification domain.

\textbf{2.} We operationalize SDI computation as a quadratic program, which can be solved efficiently.

\textbf{3.} We show that the proposed method outperforms state of the art methods on a standard e-commerce search benchmark.
 
 Next, we motivate our method and place it in the context of related prior work.

\section{ Related Work}
In this part, we emphasize the importance of evenness, brief the major baseline method $--$ DPPs, and introduce SDI as a diversity model in multi-aspect search problems.

\subsection{ Richness and Evenness}
Prior research on search diversification focused on the richness of the results
 \cite{simpson1949measurement}. Richness quantifies the number of different types the dataset of interest contains, while evenness considers the uniformity of the distribution of these types. 
However, {\em evenness} is also important for search diversification. Importantly for the e-commerce setting, it can affect users' online shopping experience. For example, consider a product catalog with 1000 A items, 500 B items, and 500 C items. Suppose, we attempt to retrieve 12 results as diverse as possible to fit on the first page of the results. Without considering evenness, the expected search results might be 6 A items, 3 B items, and 3 C items. While considering evenness, the expected results are all 4. The {\em richness} of the two versions of the results is identical. However, the information entropy of the second set of results is higher, meaning that higher {\em evenness} could increase the information that the users receive in a limited space. In multi-aspect search, evenness is even more important. In multi-aspect search, the search results are the combination of many independent single-aspect searches.
In prior research, \cite{simpson1949measurement} shows that the combinations of the even results are the most diverse. Therefore, evenness affects the richness of multi-aspect search as well.

The above discussion shows that evenness is significant for diversity analysis. However, vast majority of the diversity models for search do not consider results' evenness. Therefore, we propose to use a model considering evenness to enrich the diversity model for search. We propose to adopt the Simpson's diversity index (SDI) method from biology for this task \cite{simpson1949measurement}. But first, we formally introduce the DPP-based algorithms for result diversification and then introduce our new SDI method.

\subsection{Determinantal Point Processes for Search Diversification}
Determinantal Point Processes (DPPs) are useful in domains where repulsive effects or diversity are important aspects to model. In recommender systems and information retrieval domains, where the discovery of documents and items are significant, DPPs can be used to diversify results. 
In search problems, DPPs are a subset selection problem. We define the search space as a set $S$, whose size is $N$, and the search results are a subset $S'$, whose size is $k$. $B_{S'}$ represents the selected items in $S'$. DPPs can be formulated as: 
\begin{center}
$X = \underset{S' \subseteq S}{\operatorname{argmax}}$ det$({B^{T}_{S'}B_{S'}}))$
\end{center}
$X$ is the volume of a parallelepiped spanned by the vectors of $B$. It is determined by the vectors’ length (relevance) and their cosine similarity (similarity). By maximizing $X$, we can obtain the best subset of $S$, considering both diversity and relevance. 

However, DPPs are a richness-greedy diversification method, because $X$ does not change when we apply linear transformations to revise the evenness of $B$. Therefore, the evenness of the obtained subset of each aspect is not promised. In multi-aspect search, the results are the combination of every aspect. As we mentioned in Section 2.1, this can lead to weaker diversity of the combination.



\subsection{Simpson's diversity index}

Simpson's Diversity Index (SDI) was originally introduced in biology to measure both the evenness and richness of a habitat \cite{simpson1949measurement}, and aims to estimate the probability that two randomly selected individuals from a sample will be the same. Apparently, if the habitat has more species (Diversity), this probability is smaller. Based on Jensen's inequality, if the total number of the animals in the habitat is a constant, this probability is also smaller when the numbers of each species are the same (Evenness).

More formally, SDI is defined as the probability that two items are chosen at random and independently from a set is the same:
\begin{center}
    $D=\frac{\sum_{i=1}^R{(n_i (n_i -1))}}{N(N-1)}$
\end{center}
Where $R$ is the number of classes. $n_i$ is the number of items of the $i$th class. $N$ is the total number of all items. Smaller $D$ means higher diversity.
SDI considers both richness and evenness, according to Section 2.1, it is more suitable to diversify search results over multi-aspect items.

\section{Diversifying Multi-aspect Search Results Using SDI}
In this section, we present a method that diversifies multi-aspect search results using Simpson's diversity index. We first present an important variation of Simpson's diversity index and then explain how we apply it to search. We conclude by constructing a quadratic program to compute SDI more efficiently.

\subsection{Diversifying Multi-aspect Search Results using Simpson's diversity index }
In multi-aspect search, items are labeled by many aspects. We first consider the SDI of an aspect.
In addition to the equation in Section 2.2, we have another way to calculate SDI.

In a set, we denote the number of items in this set as $N$. $R$ is the number of items' classes. $n_i$ is the number of items of the $i$th class.

When we randomly select two items as a pair, there are $\frac{1}{2}N(N-1)$ pairs. For each pair, they belong to the same class or not. The total number of pairs belonging to the same class is
    $\sum_{i<j\leq N}{(a_i\&a_j)}$.
Where $a_i$ is the Item $i$. If two items are the same $(a_i\&a_j)=1$, else $(a_i\&a_j)=0$.
If we also count the item itself, we have:
\begin{equation}
D(p,S')=\frac{2\sum_{i,j\in S',i\leq j}{(a_{p,i}\& a_{p,j})}}{|S'|(|S'|+1)}
\end{equation}
Where $p$ is the aspect, $P$ is the aspect set, $S'$ is the selected set, $S$ is the whole set, $|S|$ is the size of $S$. $a_{p,i}$ is the Aspect $p$ of Item $i$.

If we consider all the aspects at the same time, we have:  

\begin{center}
$H(S')=\sum_{p\in P}{\omega_p \frac{D(p,S')}{D(p,S)}}$
\end{center}

In this equation, we first normalize each aspect's SDI, because different aspects' SDI may have  different scales. Then, we weigh every aspect with $\omega_p$. If we only consider diversification, $\omega_p=1$. However, in the search problem, relevance is also very important. When the item is highly relevant to some aspects. For example, when users search 'blue shoe', it is not necessary to present other colors' shoes anymore. Therefore, the 'color' aspect should be penalized. In this paper, 
$\omega_p = 1-\frac{k_{p}-1}{|S'|-1}$,
Where $k_p$ is the number of the most common aspect of Aspect $p$ in top $|S'|$ relevant results.
For convenience, we let $\phi_p = \frac{\omega_p}{D(p,S)}$. 

This transformation of SDI helps us organize it to a more efficient form to be optimized $--$ the binary quadratic program (BQP). 

\subsection{The Binary Quadratic Program}
In this section, we transfer the SDI optimization to a BQP problem, which has thorough previous research to systematically be approximated and optimized \cite{ma2010semidefinite}\cite{clarkson2010coresets}\cite{lewin2002improved}. 
Since $H(S')$ is pairwise,  it can be computed in an easy way. Let's expand $H(S')$.
\begin{center}
    $H(S')=\sum_{i,j\in S',i\leq j}{\frac{2}{|S'|(|S'|+1)}\sum_{p\in P}{\phi_p (a_{p,i}\& a_{p,j})}}$
\end{center}

    Let $Q_{i,j}={\frac{2}{|S'|(|S'|+1)}\sum_{p\in P}{\phi_p (a_{p,i}\& a_{p,j})}}$, we can reformulate $H(S')$ as
    $H(S')=\frac{1}{2}x^T Qx$,
$x$ is a vector representing $S'$. The length of $x$ equals to $|S|$. Each entry is an item of $S$. The entry of the chosen item is 1, or it is 0. 
If we further consider the relevance, we have:
\begin{equation}
     \min_x T(S')= \min_x H(S')+R(S')= \min_x \frac{1}{2}x^T Qx+b^T x
\end{equation}
Where $b$ is the relevance vector.
Consequently, it becomes a binary quadratic program problem, whose constraints are: s.t. $ x_i\in\{0,1\}, \sum_{k=1}^{|S|}{x_k}=|S'|$, where $x_i$ is the entry of $x$.

Equation 2 is a classical BQP problem. In this paper, we first relax it to a convex optimization problem \cite{ma2010semidefinite}. Then we use the Frank–Wolfe algorithm to optimize it \cite{clarkson2010coresets}. Finally, we use Goemans and Williamson procedure for the approximate solutions \cite{lewin2002improved}. Such a procedure guarantees that the solution is a close approximation that $T\in[T_{min} ,\frac{\pi}{2}T_{min}]$. 


\section{Experimental Setting and Results}
We first introduce the experimental datasets, including the queries generation and items' collection. Then, we describe the evaluation methods, initial relevance ranking, and the baseline methods used to report the experimental results. 

\subsection{Datasets}
We experiment with two datasets: a publicly available Kaggle benchmark for the searching items dataset, and a complementary dataset constructed using publicly available Amazon Web Search engine interface for the queries dataset. 

\textbf{Kaggle Shoes Price Competition Dataset (Kaggle)}:
Kaggle shoe price dataset is a list of shoes and their associated information, containing 19046 women shoes and 19387 men shoes. However, the items of this dataset also contain clothes, gloves, jewelry, and so on. Some items do not have features. To focus on the diversity problem, we remove items without features, clothes items, jewelry items, gloves items, trousers items, and toys items. At last, we have 14609 items. Besides, even though the items have many aspects, they do not have the values of many aspects.  Therefore, we only choose  21 most commonly used features: 'Season', 'Material', 'Gender', 'Shoe Size', 'Color', 'Brand',
       'Age Group', 'Heel Height', 'Fabric Material', 'Shoe Width', 'Occasion',
       'Shoe Category', 'Casual \& Dress Shoe Style', 'Shoe Closure',
       'Assembled Product Dimensions (L x W x H)', 'Fabric Content',
       'Shipping Weight (in pounds)', 'prices.offer', 'prices.amountMin',
       'prices.amountMax' and 'prices.isSale'.

\textbf{Amazon Queries Suggestions Dataset (AQS)}: 
To complement the Kaggle queries dataset, we created a new dataset automatically constructed for the ``shoes'' category using the query suggestions provided by the Amazon Web Search engine. Specifically, we collected the queries suggestions for the top-level category 'shoes'. This process resulted in 384 queries, such as 'fur lined winter coat women'.
The AQS dataset is available at the URL \url{https://docs.google.com/spreadsheets/d/17DMU5pMSiNxi05yu5pdEvrUf3b0mIAF1rYn2ypruS7A/edit?usp=sharing}.

\subsection{Baselines and Experimental Setting}
\subsubsection{Setting}
In the experiments, we search for queries and every model returns the top 10 results. We used the averaged results of all queries as the data points.

All the diversification methods considered, and our proposed method, require the relevance vectors and the similarity matrices. In this paper, BM25\cite{robertson2009probabilistic} score of each item with respect to the query.
For the item-item similarity matrix, the number of identical aspects between the items is used, which could be improved in future work.

To trade off the relevance and diversity, we introduce a parameter $\theta$. $\theta$ adjusts the relation between diversity and relevance in the following way:$(1-\theta)H(S')+\theta R(S')$.
For the proposed method, the approximate rate $\epsilon$ is 0.0001.

\subsubsection{Baselines}
We use three baselines, including Greedy version DPPs and two recent and influential variations.
\textbf{1) DPPs Greedy:}
Greedily selecting items to maximize the determinant \cite{borodin2009determinantal}.
\textbf{2) k-DPPs:}
It is a sample-based DPPs, which can accelerate the Greedy DPPs. However, since it localizes the optimization, the performance may be unstable \cite{kulesza2011k}. 
\textbf{3) Fast MAP DPPs:}
A novel approach improves the computation complexity of the maximum a posteriori (MAP) inference DPPs, which is the state of art DPPs algorithm \cite{chen2018fast}.

\subsection{Evaluation Metrics}
We now summarize the metrics used to compare the methods. 

    \textbf{Coverage Rate(CR):} The coverage rate measures the diversity of the subset \cite{castells2011novelty}. For each aspect of the item, they have at most 10 features or the number of this aspect's features in this set. The coverage rate is a diversity metric.
\begin{center}
    $CR(S')=\frac{1}{|P|}\sum_{p\in P}{\frac{|A'_p|}{\min(|S'|,|A_p|)}}$
\end{center}
Where $S'$ is the selected set, $P$ is the set of all aspects, $A'_p$ is the features of Aspect $p$ in the selected set and $A_p$ is the features of Aspect $p$ .  

    \textbf{NDCG@10:} We use normalized discounted cumulative gain (NDCG) to measure ranking results of the relevance. Since both results of DPPs and the proposed method do not have the order, we rank the selected subset based on the items' relevance before computing the NDCG@10 \cite{castells2011novelty}.
    
    \textbf{ $\alpha$-NDCG:} A variations of NDCG, which can measure both relevance and diversity \cite{castells2011novelty}.  
    
    \textbf{Variance:} We use averaged variance of every aspect's number of features to measure the evenness \cite{ricotta2017beta}.


\subsection{Results}

The results of diversity, evenness, and relevance are presented in Figure 1. The result of the computing time is illustrated in Table 1. In Figure 1, the trade-off parameter $\theta$ ranges from 0 to 1, whose step size is 0.1. For CR(a), NDCG@10(b), and $\alpha$-NDCG(c), the larger their value, the better the performance of the model is. For variance(d), a lower line means the model is evener. In Table 1, less computing time means the model is faster. Generally, the proposed method outperforms the baselines in terms of relevance, diversity, and evenness, especially when $\theta$ ranges from 0.3 to 0.7. The improvement is up to more than 15\%. In terms of computing time, the proposed method spends less time than most of the baselines.

We use CR to measure the richness of the multi-aspect results in Figure 1 (a), which shows the proposed method outperforms the baselines when $\theta$ ranges from 0.2 to 0.8. In terms of relevance, the NDCG@10 scores from Figure 1 (b) show that the proposed method has closed performances. We further use $\alpha$-NDCG to measure the relevance and diversity at the same time in Figure 1 (c), which demonstrates that the proposed approach has the best performance while $\theta$ ranges from 0.4 to 0.7. Figure 1 (d) further applies the Variance to measure the evenness of the returned results. In this figure, we find that when the model emphasizes diversity ($\theta$ closes to 0), the variance of SDI is smaller than all the baselines.
We further test the computing time of the proposed method and summarize the results in Table 1, revealing that the proposed method is slower than k-DPPs but faster than other baselines.


\begin{figure}[htb]
\centering
\[
\hspace{-0.1in}
\begin{array}{cc}
    \includegraphics[width=0.55\columnwidth]{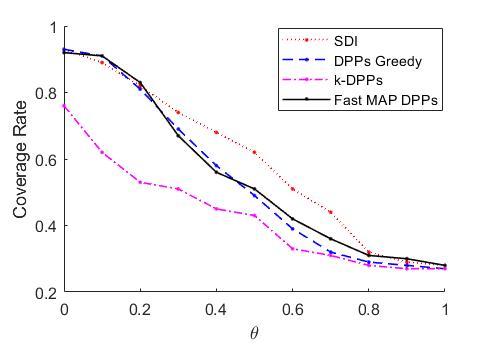}
  &\hspace{-0.3in}
 \includegraphics[width=0.55\columnwidth]{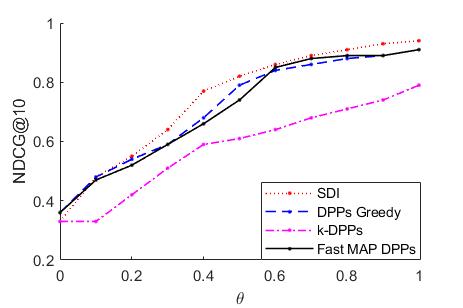}
  \\
  (a) & (b) 
  \\
  \includegraphics[width=0.55\columnwidth]{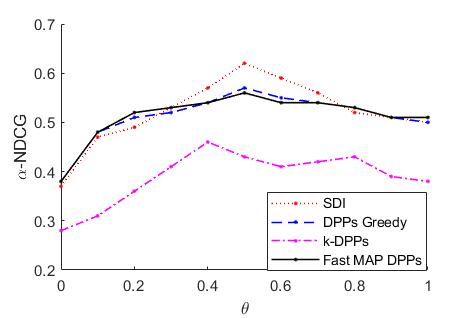}
  &\hspace{-0.3in}
 \includegraphics[width=0.55\columnwidth]{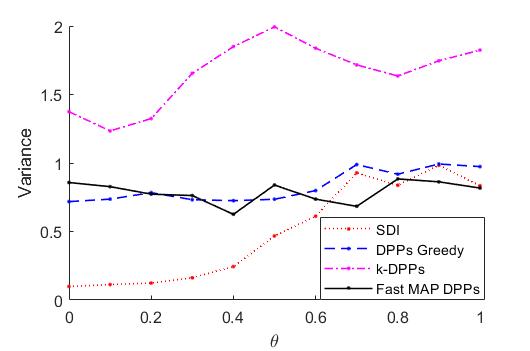}
 \\
 (c)&(d)
\end{array}
\]
\vspace{-6mm}
\caption{ Comparison of trade-off performance between relevance and diversity under different choices of trade-off parameters $\theta$ on Amazon Queries Suggestions dataset (a, c) and Kaggle Shoes Price Competition Dataset (b, d).}
\label{fig:rli-rank-results}
\end{figure}
\begin{table}
\small
  \caption{ Comparison of average running time (in milliseconds)}
  \label{tab:commands}
  \begin{tabular}{cccccl}
    \toprule
   Model& SDI&DPPs-Greedy &k-DPPs&Fast-MAP-DPPs\\
    \midrule
      Computing Time&45.32&1093.48&17.86&98.45\\
    \bottomrule
  \end{tabular}
\end{table}

\section{Analysis \& Discussion}
We now discuss the proposed method's advantages and drawbacks based on the experimental results above.
Figure 1 (a) (b) (c) show that our SDI method outperforms all the baselines while we consider diversity and relevance with similar or closed weights. Simultaneously, Figure 1 (d) shows that the evenness of SDI is substantially higher than the baselines', supporting the analysis in Section 2.

Table 1 reports the computation time of 4 methods, which aligns with their computation complexity. Greedy DPPs' is $O(N^4)$. Fast-MAP-DPPs' is closed to $O(N^3)$. k-DPPs' is $O(Nk^2)$. The proposed method has $O(1/\epsilon)$ iterations, and each iteration is just linear programming. Although the proposed method is slower than k-DPPs, k-DPPs have the worst general performance in Figure 1, which still supports the conclusion that SDI is a competitive diversity model.


\section{Conclusions}
In this paper, we presented a novel approach to multi-aspect search results diversification, by adapting the Simpson's Diversity Index (SDI) from biology for this task.  Unlike important previous SOTA family of methods for diversification, our proposed method considers both evenness and richness. Based on theoretical analysis and experimental evaluation, we show that the evenness is significant for multi-aspect search, which is common in e-commercial product search. 
By defining the relevance and diversity in different ways, SDI can be further applied to other search and recommendation methods based on learning-to-rank or neural search and recommendation algorithms \cite{zoph2016neural}. In the future, we plan to explore incorporating our SDI-based approach directly into learning-to-rank models, to further improve search and recommendation performance for e-commerce search and recommendation. 

%
\bibliographystyle{ACM-Reference-Format}
\bibliography{sample-base}

%

\end{document}